\documentclass[draft]{agujournal2019}
\usepackage{url}
\usepackage{lineno}

\usepackage{soul}
\def\dav{\color{blue}}

\draftfalse
\journalname{Geophysical Research Letters}

\usepackage{ragged2e, amsmath}
\justifying
\begin{document}

%%%%%%%%%%%%%%%%%%%%%%%%%%%%%%%%%%%%%%%%%%%%%%%
%  TITLE
%
% (A title should be specific, informative, and brief. Use
% abbreviations only if they are defined in the abstract. Titles that
% start with general keywords then specific terms are optimized in
% searches)
%
%%%%%%%%%%%%%%%%%%%%%%%%%%%%%%%%%%%%%%%%%%%%%%%

% Example: \title{This is a test title}

\title{Aperiodic clustered and periodic hexagonal vegetation spot arrays explained by inhomogeneous environments and climate trends in arid ecosystems}
\authors{David Pinto-Ramos\affil{1,2}, Marcel Gabriel Clerc\affil{2}, Abdelkader Makhoute\affil{3,4}, Mustapha Tlidi\affil{4} }

\affiliation{1}{\it Center for Advanced Systems Understanding (CASUS); Helmholtz-Zentrum Dresden-Rossendorf (HZDR), D-02826 Görlitz, Germany}
\affiliation{2}{\it Departamento de F\'isica and Millennium Institute for Research in Optics, Facultad de Ciencias F\'isicas y Matem\'aticas, Universidad de Chile, Casilla 487-3, Santiago, Chile}
\affiliation{3}{\it Facult\'e des Sciences, Universit\'e Moulay Ismail, Dynamique des Systèmes Complexes, B.P. 11201,
	Zitoune, Meknès, Morocco}
\affiliation{4}{\it Facult\'e des Sciences, Universit\'e Libre de Bruxelles (U.L.B), CP. 231, 
	Campus Plaine, B-1050 Bruxelles, Belgium}

\correspondingauthor{David Pinto-Ramos}{david.pinto@ug.uchile.cl}

\def\dav{\color{black}}

%%%%%%%%%%%%%%%%%%%%%%%%%%%%%%%%%%%%%%%%%%%%%%%
% KEY POINTS
%%%%%%%%%%%%%%%%%%%%%%%%%%%%%%%%%%%%%%%%%%%%%%%
%  List up to three key points (at least one is required)
%  Key Points summarize the main points and conclusions of the article
%  Each must be 140 characters or fewer with no special characters or punctuation and must be complete sentences

% Example:
% \begin{keypoints}
% \item	List up to three key points (at least one is required)
% \item	Key Points summarize the main points and conclusions of the article
% \item	Each must be 140 characters or fewer with no special characters or punctuation and must be complete sentences
% \end{keypoints}

\begin{keypoints}
\item Clusters of vegetation spots, lacking a Fourier characteristic mode, are an equilibrium state due to environmental heterogeneities.
\item Clustered and hexagonal patterns of spots coexist, forming a robust, {\dav universal hysteresis loop.}
\item {\dav The robustness of the hysteresis loop makes single-time dataset analysis viable, allowing for the inference of possible environmental trends.}
\end{keypoints}

%%%%%%%%%%%%%%%%%%%%%%%%%%%%%%%%%%%%%%%%%%%%%%%
%
%  ABSTRACT and PLAIN LANGUAGE SUMMARY
%
% A good Abstract will begin with a short description of the problem
% being addressed, briefly describe the new data or analyses, then
% briefly states the main conclusion(s) and how they are supported and
% uncertainties.

% The Plain Language Summary should be written for a broad audience,
% including journalists and the science-interested public, that will not have 
% a background in your field.
%
% A Plain Language Summary is required in GRL, JGR: Planets, JGR: Biogeosciences,
% JGR: Oceans, G-Cubed, Reviews of Geophysics, and JAMES.
% see http://sharingscience.agu.org/creating-plain-language-summary/)
%
%%%%%%%%%%%%%%%%%%%%%%%%%%%%%%%%%%%%%%%%%%%%%%%

%% \begin{abstract} starts the second page

\begin{abstract}{
Due to climate change, overgrazing, and deforestation, arid ecosystems are vulnerable to desertification and land degradation. As aridity increases, vegetation cover loses spatial homogeneity and self-organizes into heterogeneous vegetation patterns, a step before a catastrophic shift to bare soil. Several studies suggest that environmental inhomogeneities in time or space are crucial to understand these phenomena. Using a unified mathematical model and incorporating environmental inhomogeneities in space, we show how two branches of vegetation patterns create a hysteresis loop as the mortality level changes. In an increasing mortality scenario, one observes an equilibrium branch of high vegetation biomass that forms self-organized hexagonal-like patterns. However, when the mortality trend is reversed, one observes a branch with low biomass and no {\dav periodicity}, where vegetation spots form {\dav disordered clusters instead of a hexagonal lattice}. This behavior is supported by remote sensing and field observations and can be linked to climate change in arid ecosystems.}
\end{abstract}

\section*{Plain Language Summary}

{\dav When looking at vegetation in dry regions from planes or space, it is seen that the vegetation cover is not uniform but rather fragmented. The properties of these vegetation patches, or spots, have been suggested to warn about possible catastrophic shifts toward full vegetation death if the environmental conditions worsen. However, some of the observed patterns do not fit within the established theories. Using a model-unifying framework and accounting for a realistic environment by including changes in the biological and environmental variables from place to place, we can reproduce the unexplained patterns. Then, we show that the pattern shape is linked to the historical trends in the biological and environmental parameters. }

\section{Introduction}

The question of whether vegetation in arid climates is on the verge of collapse or not has become increasingly relevant in the context of a changing climate \cite{rietkerk2021evasion,bonachela2015termite,Martin2015,berdugo2020}. Drylands cover more than 40$\%$ of the world's territory, are a reservoir of plant diversity, and are essential for the survival of more than a third of the world's population \cite{gross2024unforeseen,mortimore2009dryland}. The focus has been put on the interactions between the individual plants with each other and with their abiotic environment to provide a robust theory for their population dynamics, predicting their behavior as the environmental parameters change over time employing mathematical models. All these models predict the emergence of spatially-periodic patterns for worsening environmental conditions \cite{thiery1995model,lefever1997origin,klausmeier1999regular,rietkerk2002self,VandeKoppel2004,borgogno2009mathematical,martinez2023integrating}. Those model-predicted patterns and their relationship with the abiotic variables are compared with actual observations, 
granting the models a good qualitative predictive power \cite{deblauwe2008global,Deblauwe2011}. Then, supported by the model predictions, vegetation patterns are regarded as an early-warning indicator of possible collapse to bare soil and, at the same time, seen as a robust distribution of the biomass resistant against external perturbations due to the theoretical multistability of states \cite{rietkerk2021evasion}. 

The idea of spatial patterns being relevant for the ecosystem is not unique to vegetation in arid climates but rather general in diverse ecosystem modeling, such as the coexistence of species \cite{hassell1994species}, savanna landscapes \cite{eigentler2020spatial}, seagrass meadows \cite{ruiz2023self}, or mussel beds \cite{van2008experimental,liu2014pattern}, to mention a few. Thus, carefully analyzing the nonlinear dynamics of such spatial patterns has attracted the attention of scholars to better understand how they affect ecosystems. In arid ecosystems, the vegetation pattern morphology is correlated with the aridity of the environment \cite{Deblauwe2011}, displaying homogeneous covers, gaps, labyrinths or stripes, and spots as the typical sequence observed from lower to higher aridities. This poses the spotted pattern as particularly relevant, being the last one in the sequence before the transition to bare soil occurs \cite{Tlidi2008,lejeune2004vegetation,rietkerk2004self,VonHardenberg2001}. This transition has been the focus of discussion, since researchers have theoretically identified either abrupt transitions leading to hysteresis loops, or continuous-like transitions \cite{rietkerk2021evasion,martinez2023integrating}; two phenomena incompatible at first glance. 

Despite models predicting the overall morphology of the pattern, the details of the spatial structure show strong disagreement between the model-predicted patterns and the observed ones, such as the distorted and irregular appearance of natural patterns compared with the rather perfect and regular (like a crystal) patterns obtained from simulations {of biomass density models} \cite{kastner2024scale,pinto2023topological,yizhaq2014effects}. Instead of being just a small perturbation affecting the looks of the spatial pattern, it has been shown that the observed distortion of the patterns could be a symptom of complex dynamics triggered by noise \cite{d2006vegetation}, heterogeneity in the water soil diffusion \cite{yizhaq2014effects} or heterogeneity in the plant mortality \cite{pinto2022vegetation}. Strikingly, including those ingredients changes the shape of the diagram of states of the average biomass for worsening environmental conditions, inducing a smooth transition instead of an abrupt collapse to bare soil \cite{yizhaq2014effects,Martin2015,yizhaq2016effects,pinto2022vegetation}. For those reasons, heterogeneity in the landscape and noise have been recognized as an important ingredient that could drive the resilience of patterned vegetation even further, with measurable impact \cite{yizhaq2017geodiversity}. On the other hand, it has been proposed that patterns with distortion and irregularity could be observed even in homogeneous and deterministic landscapes due to a family of solutions known as localized structures, well known in nonlinear dynamics. Nevertheless, this mechanism is not robust, as localized solutions exist in a narrow region of the bifurcation parameter close to the so-called Maxwell point \cite{woods1999heteroclinic,coullet2000stable,clerc2005localized,burke2007homoclinic}. {\dav Similarly, irregular patterns have been described as transient states or single snapshots of stochastic processes \cite{d2007noise,scanlon2007positive,martinez2013spatial,surendran2025spatial}; there, the patch size distribution has been linked to ecosystem health and functioning \cite{scanlon2007positive,kefi2007spatial}, and irregular patterns have been recently studied as hyperuniformly ordered \cite{torquato2018hyperuniform}, with consequences for water capture and resilience \cite{ge2023hidden}. Nevertheless, irregular patterns and their apparent stability can so far not be captured by density-based equations}.  In the case of spots---the focus throughout this work---apparent {\dav stable} random arrays of spots were first reported by \citeA{lejeune2002localized}. However, employing a nonlinear analysis of the interaction between localized structures,  \citeA{berrios2020repulsive} showed that spots' (on top of bare soil) interactions are always repulsive, making it impossible to form nonperiodic, irregular, or distorted patterns. That result highlights the need for an alternative, robust mechanism that produces {\dav stable} irregular arrays of vegetation spots, opening the questions of what the ecological consequences of observing irregular arrays of spots are and how they compare to the well-known regular hexagonal arrays of spots. 

In this work, we aim to explain a mechanism that allows for the formation of {\dav stable} irregular (nonperiodic) and regular (periodic) spotted patterns within a unified theoretical model. We argue that a combination of a heterogeneous environment (space-varying parameters) and a trend in time for the environmental adversity (for example, an increasing or decreasing aridity level over decades) could explain the formation of hexagonal-like arrays of spots and irregular arrays of spots. We call the last ones \textit{clusters} of spots due to the self-replicating process that forms them in privileged portions of space \cite{bordeu2016self,tlidi2018extended}. Our theory shows that hexagons and clusters of spots are the two branches of a hysteresis loop, similar to what was proposed by \citeA{yizhaq2014effects} and \citeA{yizhaq2016effects} for the average biomass. However, we identify the branches as two different morphological spotted patterns easily distinguishable in their Fourier spectrum. These two branches---clustered and hexagonal patterned spots---occur independently of the model framework and for broad portions of the parameter space. This not only offers a robust explanation for the observation of irregular arrays of vegetation spots, but also reconciles gradual transitions with hysteresis loops, both being a consequence of environmental heterogeneity promoting additional equilibrium states. Moreover, it suggests that spotted pattern morphology serves as an indicator of environmental adversity trends, which can be inferred from single-time data analysis. We start by presenting remote-sensing and field observations of two similar plants exhibiting a spotted pattern and similar aridity and individual structural properties. Interestingly, their spatial organization properties show differences, allowing us to classify them as either hexagons or clusters. Then, we present a theoretical model that unifies the two main families of deterministic, continuous in space and time models of vegetation population dynamics in arid environments: interaction-redistribution and water-biomass models. Introducing spatial heterogeneity in the parameters and computing the diagram of equilibrium states, we unveil the hysteresis loop between hexagonal and clustered patterns. The unified model employed makes our results model-independent. Moreover, a four-dimensional parameter sweep shows the robustness of our results, which hold for vast combinations of parameters and independently of the bifurcation structure of the homogeneous or patterned solutions. Lastly, we measure the change in aridity in both plant environments, which coincides with our theoretical predictions. Our results suggest that environmental heterogeneity could be a fundamental ingredient to better model the population dynamics of vegetation in arid environments. We conjecture that the spotted pattern classification (hexagonal or clustered) could be an indicator of bettering or worsening conditions in the ecosystem.

\section{Results}
\subsection{Remote-sensing and field observations}

\begin{figure}[t]
\centering
\includegraphics[width=0.95\textwidth]{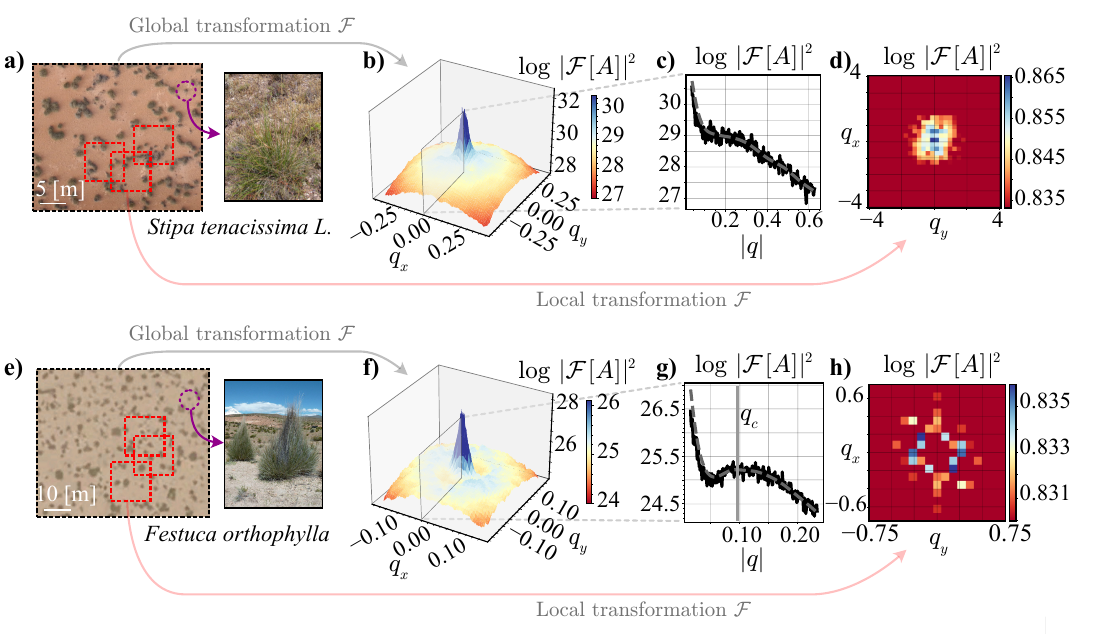}
\caption{Clustered and hexagonal arrays of vegetation patches analysis. a)--d) Morocco, Enjil region (Boulmane Province) $18^{o}$ $49.3^{'}$S, $35^{o}$ $37.4^{'}$E., 
		patches formed by alpha or {\it Stipa tenacissima L.}. 
		e)--h) Argentina (Catamarca, northwestern Argentina plateau) $23^{o}$ $25.8^{'}$S, $66^{o}$ $4.4^{'}$W. 
		Patches formed by paja brava or {\it Fetusca orthophylla} plants. 
		Panels b) and f) illustrate the 2D Fourier spectrum of the binarized vegetation cover field, $A$, which is obtained by taking the Fourier transform to the whole pattern image. Panels c) and g) 
		depict the mean profile of the 2D Fourier spectrum in the radial direction. Panels d) and h) show
		the windowed Fourier spectrum of each cover, obtained by taking a local Fourier transform of several randomly picked windows of the whole pattern, aligning them along the principal direction, and averaging them.}
\label{F1}
\end{figure}

We focus our study on two different plants with remarkable similarities that make them candidates to exhibit self-organized spatial patterns: \textit{Stipa tenacissima L.} (Fig. \ref{F1} a)--d)), natural from the west Mediterranean \cite{tlidi2018observation}, and \textit{Festuca orthophylla} (Fig. \ref{F1} e)--h)) from west South America \cite{couteron2014plant}. Both of these species thrive in regions with aridity levels as high as $\textbf{Ar}\sim 0.65$, where $\textbf{Ar}=1- P/PET_0$, and $P$ is the mean annual precipitation and $PET_0$ the potential evapotranspiration. Furthermore, their structural ratio, $\epsilon = L_f / L_c$ is of similar magnitude, being $\epsilon \sim 0.25$. The structural ratio corresponds to the quotient of the facilitative to competitive radii \cite{lefever2009deeply,Tlidi2008}. For \textit{Festuca orthophylla} we used the values reported by \citeA{couteron2014plant}, where $L_c \sim 40 cm$ is estimated with the lateral extension of the roots and $L_f \sim 10 cm$ is estimated as half the height of the ramets. Employing the same estimators for \textit{Stipa tenacissima L.}, we get values of $L_c \sim 170 cm$ and $L_f \sim 40 cm$. Having the same values of the aridity level and the structural ratio, they are likely (up to the facilitative to competitive interaction strength quotient, which for \textit{Festuca orthophylla} is estimated around $\sim 1.1$ by \citeA{couteron2014plant} ) to form patterns in the model of \citeA{lefever2009deeply} and \citeA{Tlidi2008}. 

Species \textit{Stipa tenacissima L.} and \textit{Festuca orthophylla} are observed to arrange in spots of vegetation surrounded by bare soil, namely, a spotted pattern, as seen in Fig. \ref{F1} a) and e). These spots live in a complex topography, where a gentle slope {\dav with small variations --although not steep enough to induce anisotropic patterns \cite{Deblauwe2011}--can be noted}. When analyzing the spatial pattern in detail, one finds fundamental differences. Unexpectedly, spots of \textit{Stipa tenacissima L.} show an arrangement in space that does not have a dominant wavelength. This is seen in the Fourier power spectrum for the spatial modes of the binarized aerial picture of the pattern (the field $A$), depicted in Fig. \ref{F1} b). Performing the radial average, one can observe a monotonically decaying profile, as seen in Fig. \ref{F1} c). Furthermore, analyzing the pattern locally with the objective of identifying its structure \cite{echeverria2023effect,echeverria2020labyrinthine}, one finds no structure indicative of a cell that repeats in space forming a pattern, as seen in Fig.~\ref{F1} d). On the other hand, spots of \textit{Festuca orthophylla} exhibit the characteristics of a hexagonal pattern with a degree of disorder. One identifies this by computing the Fourier power spectrum, where one observes local maxima at different wavelengths, as seen in Fig.~\ref{F1}~f). At the global scale (the whole pattern), the hexagonal structure is not clear, but a dominant length is evident, as shown in the radially averaged power spectrum seen in Fig.~\ref{F1} g). Analyzing the pattern locally, one finds the hexagonal structure in the local Fourier power spectrum, indicative of a hexagonal cell that repeats in space, as seen in Fig.~\ref{F1} h). However, different directions of the hexagonal cell make the global power spectrum radially symmetric. Thus, the spotted pattern of \textit{Festuca orthophylla} corresponds to a disordered hexagonal pattern. The patterns observed for \textit{Stipa tenacissima L.} and \textit{Festuca orthophylla} can not be reproduced with the current models of vegetation population dynamics that consider a homogeneous environment. However, a heterogeneous environment could be a candidate to explain the phenomena observed, considering that it is capable of distorting the patterns obtained in models with homogeneous parameters \cite{yizhaq2014effects,pinto2022vegetation}.

\subsection{Theoretical modeling}

Mathematical models for the vegetation spatiotemporal population dynamics exist in several forms and perspectives. The first ones correspond to stochastic models for individuals \cite{thiery1995model,bolker1999spatial,law2000dynamical}. However, due to their complexity and the difficulty of generating analytical predictions from them, several models based on continuous time and space have arisen. Continuous models deal with biomass density instead of individuals and are described by deterministic partial differential equations, which favor analytical treatment. From this family, there exist two types of models: Interaction redistribution models based on a single equation and nonlocal interactions of the biomass field first proposed by \citeA{lefever1997origin}, and coupled water-biomass density equations formulated by \citeA{klausmeier1999regular}. Inspired by these pioneering works, several model adaptations and generalizations have been developed \cite{lejeune1999model,VonHardenberg2001,hillerislambers2001vegetation,rietkerk2002self, gilad2004ecosystem,hernandez2004clustering,Tlidi2008,lefever2009deeply,martines2013vegetation,ruiz2017fairy,tlidi2024non}, see references \cite{borgogno2009mathematical,martinez2023integrating} for a comprehensive review. These approaches have been validated independently, being parameterized with in-site measurements \cite{barbier2008spatial,couteron2014plant} or estimates from the literature \cite{meron2015ModelNew}. Then, we decided to employ a reduced model that can be derived from both the interaction redistribution model and the water-biomass model, making our results independent of the modelling approach. The reduction is based on a change of variables to the center manifold near a codimension-4 point in parameter space where the bare soil state changes its stability, bistability between the bare soil and a uniform populated state emerges, and a Turing bifurcation in the limit of vanishing wavenumber occurs. For a detailed derivation and the description of each parameter in terms of ecologically relevant quantities, see the Supplementary Material. The equation reads \cite{Tlidi2008}
\begin{eqnarray}
\partial_t b = (-\eta + \kappa b - b^2)b + (d-\gamma b)\nabla^2b -  \alpha b \nabla^4 b,
\label{eq1}
\end{eqnarray}
where $b$ is the biomass density field as a function of space $\mathbf{r}$ and time $t$. $\partial_t$ denotes the  partial derivative with respect to $t$, $\nabla^2=\partial_{x}^2+\partial_y^2$ is the laplacian operator in 2D, and $\nabla^4 = (\nabla^2)^2$ is the bilaplacian operator in 2D.  $\eta$ is a proxy for the effective linear death rate (that is, birth rate minus death rate) of the population, which is affected by external factors such as grazing or precipitation. $\kappa$ is a proxy for the cooperativity of the species, that is, $\kappa>0$ allows survival past the bifurcation at $\eta=0$. $d$ is the dispersal of the population in the diffusive approximation. $\gamma$ is a proxy for the negative feedback scale that induces patterns. $\alpha$ is a spatial scale that ensures the saturation of the model (that is, the solutions do not diverge). Equation \eqref{eq1} is complemented with boundary conditions, 
which we consider, for the sake of simplicity, to be periodic throughout this study.

To include the effect of a heterogeneous environment, we promote the parameters of Eq.~\eqref{eq1} to be functions of space. Then, the parameters determining the spatiotemporal evolution of the biomass density field are affected by the local environment due to, for example, topography \cite{gandhi2018topographic}, geodiversity of the soil \cite{yizhaq2017geodiversity}, a variable soil-water diffusivity \cite{yizhaq2014effects}, or various factors driving mortality \cite{pinto2022vegetation,echeverria2023effect}, to mention a few. 
We will focus on a single parameter due to the high complexity of studying all of them simultaneously. For this reason, the linear growth of the population $\eta$ is promoted to a function of space $\eta \rightarrow \eta(\mathbf{r}) = \eta + \Gamma \xi(\mathbf{r})$, {\dav where $\Gamma$ describes the magnitude of the $\eta$ spatial variations}. With the purpose of modeling the environmental heterogeneity, the spatial function $\xi(\mathbf{r})$ is modeled as a Gaussian random variable with zero mean and no correlations; that is, $\langle \xi(\mathbf{r}) \rangle =0$ and $\langle \xi(0) \xi(\mathbf{r})\rangle = \delta(\mathbf{r})$, where $\delta(\mathbf{r})$ is the Dirac delta distribution. Note that different ecological parameters that are heterogeneous in nature would be translated into different parameters of Eq. \eqref{eq1} being heterogeneous. Then, the reduced equation derived provides a robust framework for analyzing heterogeneities and their relationships across different models.

\subsection{Numerical simulation predictions}

We first {\dav recall} the model Eq. \eqref{eq1} outcome when considering a homogeneous environment with the objective of highlighting the differences between spatial patterns in a homogeneous and a heterogeneous environment. Several authors have studied interaction redistribution and water-biomass models under homogeneous conditions, all of them finding a typical cascade of transitions as the environmental conditions get worse. These transitions correspond to homogeneous cover, gapped periodic pattern, labyrinthine pattern, spotted periodic pattern, and finally, bare soil \cite{lejeune2004vegetation,lejeune2002localized,rietkerk2002self,gilad2004ecosystem,VonHardenberg2001,borgogno2009mathematical}. In most of these studies, an abrupt transition follows the spotted pattern, collapsing to the bare soil state discontinuously. Nevertheless, some studies suggest that a continuous-like transition is possible \cite{martines2013vegetation,martinez2023integrating} with spots decreasing in amplitude smoothly down to the bare soil state (although this has been only shown numerically), and others suggest that localized structures could mediate a stepped transition to bare soil, arguing that patterns with a variable number of spots can smooth the transition \cite{zelnik2013regime,zelnik2018regime,rietkerk2021evasion}.  However, these studies do not address the rather general observation of irregular patterns and their possible connection with heterogeneous environments.

\begin{figure}[h!]
\centering
\includegraphics[width=0.8\textwidth]{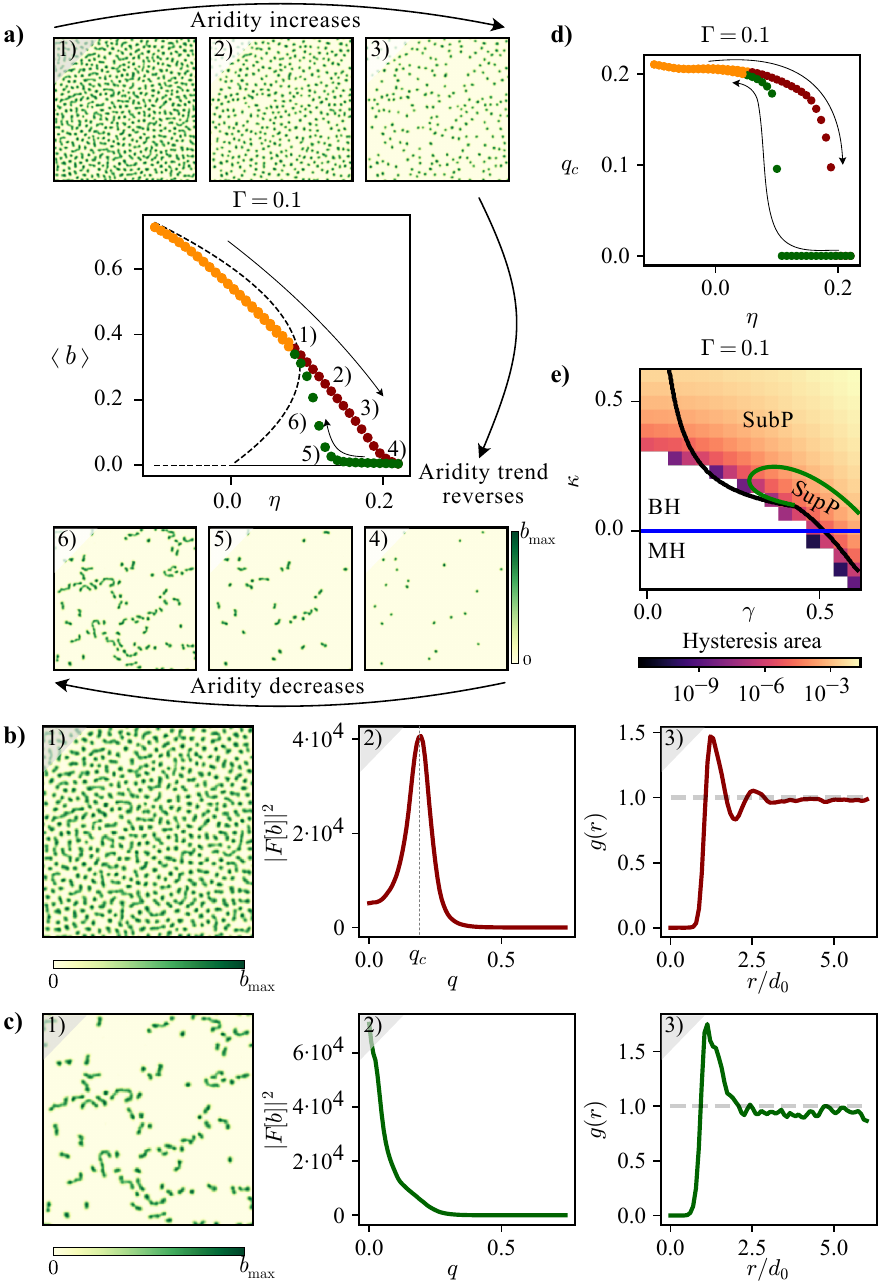}
\caption{Numerical simulation results of Eq. \eqref{eq1} in a heterogeneous environment. a) diagram of states with examples of the vegetation patterns numbered from 1) to 6). Continuous (dashed) lines show stable (unstable) homogeneous equilibria.  Dots are obtained numerically. The color highlights the pattern morphology.  b) analysis performed on the upper branch of solutions, exhibiting the characteristics of a periodic pattern with disorder. Inset $1$ shows an example of the pattern, $2$ depicts the radially averaged Fourier power spectrum with the characteristic wavenumber highlighted, and $3$ corresponds to the pair correlation function of the spots' positions. c) same analysis as b), but for the lower branch of solutions, exhibiting the characteristics of a non-periodic pattern with clustering properties. d) shows the diagram of states for the characteristic wavenumber versus $\eta$.  e) area of the hysteresis loop shown in panel a) for different $(\gamma, \kappa)$ combinations. The blue curve represents the nascent bistability that separates the monostable (MH) and bistable (BH) regimes for the homogeneous solutions. The black curve is the Turing instability. The green curve separates the region of supercritical (SupP) and subcritical (SubP) patterns. Simulation parameters are $\kappa=0.6$, $d=0.02$, $\gamma=0.5$, $\alpha=0.125$, $\Gamma=0.1$, on a grid with $192^2$ elements with a space discretization $dx=2/3$.
		}
\label{F2}
\end{figure}

In a heterogeneous environment, the structures that model Eq. \eqref{eq1} forms at equilibrium are different compared to a homogeneous environment. We measured the expected outcome of a heterogeneous environment by performing numerical simulations with 18 different realizations of the inhomogeneity function $\xi(\mathbf{r})$. In addition, we defined the equilibrium by a tolerance criterium, ensuring that $|\partial_t b (\mathbf{r}, t)| < Th$ for every point $\mathbf{r}$, with $Th$ a threshold value that we set at $10^{-6}$ (see the Supplementary Materials for details on the sensitivity of this threshold). Increasing the environmental adversity (modeled by increasing the mortality $\eta$), one observes that no abrupt jumps between pattern morphologies and the bare soil occur, in line with previous predictions analyzing the average biomass \cite{Martin2015,yizhaq2016effects,pinto2022vegetation}. This is seen in Fig. \ref{F2}. Panel a) shows the diagram of states for the average biomass, where one sees that for increasing mortality, the system remains in a branch of states (insets $1$, $2$, and $3$) with the characteristics of a disordered periodic pattern. If one stops increasing the mortality before all the spots die, as in inset $4$ of panel a), and the mortality starts to decrease instead, one sees that the equilibrium state follows a different branch of solutions depicted by insets $5$ and $6$ of panel a). This corresponds to the hysteresis loop induced by the heterogeneous environment \cite{yizhaq2014effects}, which creates branches of higher biomass and branches of lower biomass depending on the history of the control parameter. However, not only are the two branches of solutions differentiated by the average biomass, but they can also be differentiated in their pattern morphology. This is highlighted in panels b) and c), showing the typical mosaics of the upper (b) and lower (c) branch seen in panel a) in inset $1$, the Fourier power spectrum in inset $2$, and the pair correlation function of the spots' positions in inset $3$. One can see that a characteristic wavenumber exists only in the upper branch, which also exhibits a damped oscillating pair correlation function for the spots' positions. Thus, we classify this pattern as a disordered hexagonal spotted pattern. On the other hand, the lower branch exhibits no characteristic wavenumber, and the pair correlation shows a peak followed by monotonous decay. Hence, we refer to this branch as a clustered spotted pattern. One can see how the hysteresis behavior of the average biomass is also observed for the characteristic wavenumber $q_c$, as shown in panel d). This means that decreasing or increasing adversity scenarios can be identified by analyzing the spotted pattern morphology. 

The results of our numerical simulations are not limited to the parameters used in Fig. \ref{F2} panels a)-d), but occur throughout the whole parameter space. To show this, we performed the simulation protocol (that is, quasi-statically sweeping the simulation for increasing and then decreasing the $\eta$ parameter, see more details in the Supplementary Materials) while varying three additional parameters: $\kappa$, $\gamma$, and $\Gamma$, and measured the area of the hysteresis loop. Sweeping $\kappa$ and $\gamma$ allows for exploring different bifurcations' structures in the nonlinear system, and $\Gamma$ allows for assessing different heterogeneity intensities. The results for $\Gamma=0.1$ are shown in panel e). One can see that a non-vanishing hysteresis loop area is observed for a broad portion of the parameter space. Moreover, it occurs at any bifurcation structure (for the $\eta$ parameter) considered, represented by the regions enclosed by colored curves. In particular, the blue curve represents the onset of bistability between homogeneous states: MH (BH) corresponds to the region of monostable (bistable) homogeneous solutions. The black curve represents the Turing instability (patterns to the right of it). These patterns bifurcate from the homogeneous solutions either subcritically (SubP) or supercritically (SupP), behaviors which are separated by the green curve in the $(\kappa, \gamma)$ space. Interestingly, a non-vanishing hysteresis loop area persists in any of the aforementioned regions. Furthermore, the situation does not qualitatively change by varying the $\Gamma$ parameter up to values as high as $\Gamma\sim0.27$ (see the Supplementary Materials).

\section{Discussion and conclusions}

The hysteresis loop between two different types of spotted patterns is a robust characteristic of vegetation patterns induced by environmental heterogeneity. This is due to the generality of the equation we employed in our numerical simulations, and, more importantly, because it is a phenomenon observed for a broad range of parameters. Indeed, as shown in Fig. \ref{F2}, the results hold for most combinations of parameters that produce patterns in a corresponding homogeneous environment. Interestingly, the hysteresis loop persists even when the homogeneous solutions exhibit monostability. Although hysteresis loops can occur in any regime of the system considered (MH, BH, SubP, or SupP), their area (and thus, the difference between the upper and lower branches) increases as the $\gamma$ or $\kappa$ parameter increases, highlighting its nonlinear nature. We remark that the difference between the upper and lower branches in non-pattern-forming regions becomes more subtle in Fourier space; however, this issue is outside the scope of this work, as it is a phenomenon not involving spotted patterns.

The two branches of solutions that emerge due to heterogeneity have an evident difference in total biomass. We attribute this to the pinning-depinning phenomenon of the pattern-invasion front interacting with the environmental heterogeneity. The mechanism is that, despite the favored nucleation of spots in privileged portions of space, they cannot propagate the biomass further---for example via self-replication \cite{tlidi2018extended,bordeu2016self} or front propagation of the patterned state--- due to a strong Peierls-Navarro barrier for the front motion exacerbated due to the heterogeneity \cite{castillo2020frontpinning}. This means that in the lower branch, fewer spots will be created, and thus, less biomass is observed. This mechanism creates a high correlation between the positions of the spots at short distances, a phenomenon we call \textit{clustering}, as clusters of spots are formed around the regions favored by the heterogeneity. This contrasts with the case of a homogeneous environment, where the propagation front is not stopped and all space is colonized, leaving a regular pattern of spots at equilibrium. An example of these phenomena is shown in the Supplementary Materials and the Supplementary Animations. Furthermore, this mechanism induces a morphological difference between the upper and lower branches of solutions. The Fourier power spectrum acquires a different qualitative shape. It is non-monotonic for the upper branch exhibiting a characteristic wavenumber, but is monotonically decaying in the lower branch (or clustering branch) and no characteristic wavenumber is observed (see Fig.~\ref{F2}). Such a monotonically decaying power spectrum is often characteristic of a scale-free pattern \cite{von2010periodic}; however, in the clustering branch, spots can still be identified from each other and share similar properties (circularity and size), allowing to compute the pair correlation function of their positions. Our results remain valid for a substantial range of values of the heterogeneity intensity, $\Gamma$ (see the Supplementary Material). Nevertheless, we note that by increasing enough the intensity of the heterogeneities measured by the parameter $\Gamma$, spots would stop being recognizable as such, and a true scale-free pattern would emerge; this is similar to what is observed in the work of \citeA{echeverria2023effect}, who notes that a scale-free pattern could emerge either by increasing the correlation length of the heterogeneity mask as well as the heterogeneity intensity. Thus, the clustering branch of solutions emerges as a state between a regular periodic pattern and a scale-free pattern mediated by the heterogeneity properties. {\dav One limitation of our model is the rather ideal spatial heterogeneity used, in reality, heterogeneities can be more complex, such as the ones induced by topography or soil content. As a further perspective, modeling heterogeneities with data-derived functions would provide a refinement of our theory. Additionally, considering that our theory produces stable irregular patterns, recent advancements such as the hidden order of vegetation patterns characterized by hyperuniform distributions \cite{ge2023hidden} could be studied in this model, which could unveil additional properties and functions of either the disordered hexagonal or clustered spotted patterns found here.}

\begin{figure}[t!]
\centering
\includegraphics[width=0.95\textwidth]{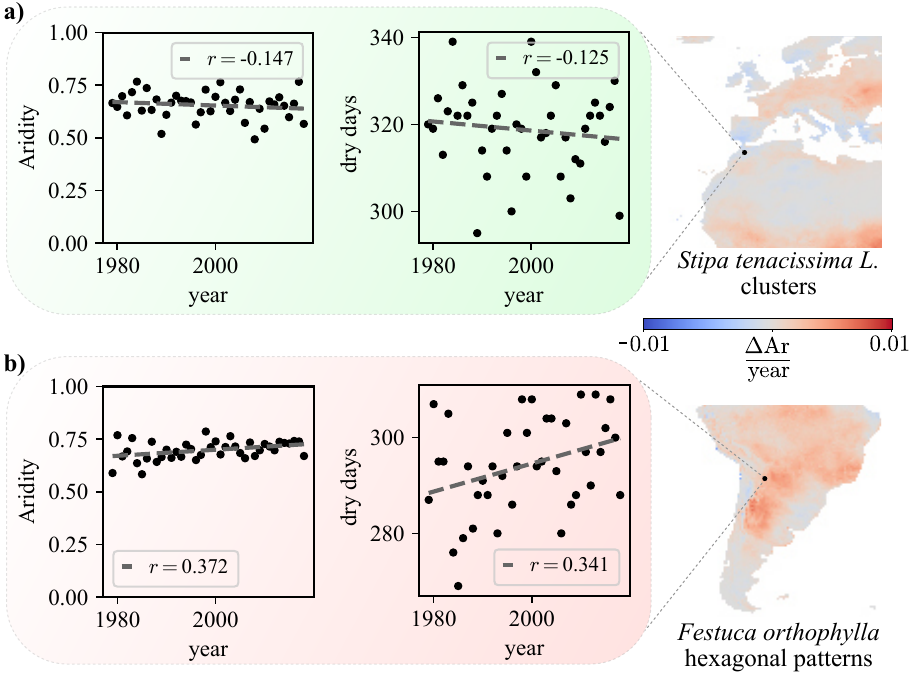}
\caption{Change rate in the aridity and yearly dry days for the vegetation's environments shown in Fig. \ref{F1}. Data cover is from 1979 to 2018. Panel a) shows the bioclimatic variables for the analyzed vegetation pattern in Morocco. Panel b) illustrates the same for the analyzed vegetation pattern in Argentina. The right-most panels are a map indicating the geographical region of the vegetation pattern, and show the aridity trend in the surroundings. Blue (red) regions show the landscapes in which the aridity has decreased (increased) from 1979 to 2018.}
\label{F3}
\end{figure}

The theoretical relationship between the morphology of the spotted pattern, being either clustered or hexagonal, and the history of the effective linear death rate parameter $\eta$ could be of interest. This property brings the hypothesis that by just analyzing a picture of the vegetation cover distribution, one could deduce if the environmental conditions are getting worse or better. By environmental conditions, we refer to any quantity that affects the population's birth or death rates and, in the models used, could correspond to the aridity, precipitation, or evaporation rate. This would allow one to identify recovering or degrading ecosystems in the sense that an increase or decrease in biomass density is expected. To explore this possibility, we analyzed the behavior of the aridity parameter and the number of yearly dry days (precipitation $< 2$ mm) in the two ecosystems observed, as illustrated in Fig. \ref{F3}. The data is obtained from the public database by \citeA{copernicusData2021}. These variables are deemed relevant due to the fast reaction of the environment and the plants \textit{Festuca orthopylla} and \textit{Stipa tenacissima L.} to rain events \cite{pugnaire1996response,monteiro2011functional}. Interestingly, \textit{Stipa tenacissima L.} patterns with a clustered morphology have lived through a period of decreasing aridity (Pearson $r=-0.147$, $p=0.362$) and yearly dry days (Pearson $r=-0.125$, $p=0.441$), as shown in Fig. \ref{F3} a). On the other hand, \textit{Festuca orthopylla} patterns with a hexagonal morphology have lived through a period of increasing aridity (Pearson $r=0.372$, $p=0.018$) and yearly dry days (Pearson $r=0.341$, $p=0.030$), illustrated in Fig. \ref{F3} b). Although these results agree with the theoretically predicted relationship between the morphology and the parameter history, more studies are needed to validate the use of it as an ecosystem health indicator worldwide. In addition, it is important to note that despite the parameter history being a direct cause of the pattern morphology, it is not a unique one. Another possible mechanism to produce such morphologies is to have fixed environmental conditions but start with different initial conditions. An initial condition composed of sparse spots and low biomass would converge to the clustered branch, but an initial condition composed of highly packed spots and high biomass would converge to the hexagonal branch for the same parameter values. Thus, our theory suggests that knowledge of the initial conditions of the vegetation is equally important (compared to the history of the environmental variables) to draw any conclusions about their future. 

To conclude, we have shown that environmental heterogeneities are a plausible explanation for the observation of disordered and irregular arrays of spots, which can otherwise not be observed in models employing homogeneous environmental conditions. We can classify spotted patterns as either disordered-hexagonal or clustered, and we illustrate that a hysteresis loop between clustered and hexagonal patterns of vegetation spots is a robust response of arid ecosystems subjected to environmental heterogeneity. This result connects the spotted pattern morphology and spatial distribution properties with environmental trends, such as the increasing or decreasing of aridity over time. Within our theory, spotted patterns in a heterogeneous environment are expected to always transition smoothly towards the bare soil state as environmental conditions worsen, and a slow recuperation through a clustering solution branch is expected once the environmental trend reverses. The robustness of this phenomenon is supported by the general equation employed, which describes two families of models, and the substantial numerical simulations performed over a broad region of the non-dimensional parameter space. Hence, our work highlights the importance of modeling heterogeneous environments in studying vegetation pattern formation and their relation to bioclimatic variables trends and the ecosystem response to them.

\section*{Open Research Section}
Data on aridity and yearly dry days is obtained from \citeA{copernicusData2021}. Simulation data, observational data, simulation protocol codes, and data analysis codes are publicly available in the Rossendorf Data Repository (RODARE) \cite{pinto2025_3983}. 

\section*{Conflicts of Interest}
The authors declare that they have no competing interests in this study.

\acknowledgments

D.P.-R. acknowledges the financial support of ANID National Ph.D. scholarship 2020-21201484. 
M.G.C. acknowledges the financial support of ANID-Millennium Science Initiative Program-ICN17$\_$012 (MIRO) and FONDECYT project 1210353. 
This work was partially funded by the Center of Advanced Systems Understanding (CASUS), 
which is financed by Germany’s Federal Ministry of Education and Research (BMBF) 
and by the Saxon Ministry for Science, Culture, and Tourism (SMWK) with tax funds on the basis of the budget approved by the Saxon State Parliament.
M.T. is a Research Director at Fonds de la Recherche Scientifique FNRS. We would also like to thank the CNRST for its support under the FINCOME programme (N L71/2022). The authors gratefully acknowledge the financial support of Wallonie Bruxelles
International (WBI). 

\bibliography{Vegetation_curated}

\end{document}